\documentclass{aastex62}

\usepackage[utf8]{inputenc}

\usepackage{graphicx}
\usepackage{amsmath}

\usepackage{gensymb}

\newcommand{\ucb}{\affiliation{Physics Department, University of California, Berkeley, CA 94720-7300, USA}}
\newcommand{\ssl}{\affiliation{Space Sciences Laboratory, University of California, Berkeley, CA 94720-7450, USA}}

\begin{document}

\title{Density And Velocity Fluctuations of Alpha Particles in Magnetic Switchbacks}

\author[0000-0001-6077-4145]{Michael D. McManus}\ucb\ssl
\correspondingauthor{Michael D. McManus}
\email{mdmcmanus@berkeley.edu}
\author[0000-0003-1138-652X]{Jaye Verniero}
\affil{Heliophysics Science Division, NASA, Goddard Space Flight Center, Greenbelt, MD 20771, USA}
\author[0000-0002-1989-3596]{Stuart D. Bale}\ucb\ssl
\author[0000-0002-4625-3332]{Trevor A. Bowen}\ssl
\author[0000-0001-5030-6030]{Davin Larson}\ssl
\author[0000-0002-7077-930X]{J. C. Kasper}
\affiliation{Climate and Space Sciences and Engineering, University of Michigan, Ann Arbor, MI 48109, USA}
\affiliation{Smithsonian Astrophysical Observatory, Cambridge, MA 02138 USA}
\author[0000-0002-0396-0547]{Roberto Livi}\ssl
\author[0000-0002-6276-7771]{Lorenzo Matteini}
\affil{The Blackett Laboratory, Imperial College London, London, SW7 2AZ, UK}
\author[0000-0003-0519-6498]{Ali Rahmati}\ssl
\author[0000-0002-4559-2199]{Orlando Romeo}\ssl
\author[0000-0002-7287-5098]{Phyllis Whittlesey}\ssl
\author[0000-0002-9202-619X]{Thomas Woolley}
\affil{The Blackett Laboratory, Imperial College London, London, SW7 2AZ, UK}

\begin{abstract}
    Magnetic switchbacks, or sudden reversals in the magnetic field's radial direction, are one of the more striking observations of Parker Solar Probe (PSP) thus far in its mission. While their precise production mechanisms are still unknown, the two main theories are via interchange reconnection events and in-situ generation. In this work density and abundance variations of alpha particles are studied inside and outside individual switchbacks. We find no consistent compositional differences in the alpha particle abundance ratio, $n_{\alpha p}$, inside vs outside, nor do we observe any signature when separating the switchbacks according to $V_{\alpha p}/V_{pw}$, the ratio of alpha-proton differential speed to the wave phase speed (speed the switchback is travelling). We argue these measurements cannot be used to rule in favour of one production mechanism over the other, due to the distance between PSP and the postulated interchange reconnection events. In addition we examine the 3D velocity fluctuations of protons and alpha particles within individual switchbacks. While switchbacks are always associated with increases in proton velocity, alpha velocities may be enhanced, unchanged, or decrease. This is due to the interplay between $V_{pw}$ and $V_{\alpha p}$, with the Alfvénic motion of the alpha particles vanishing as the difference $|V_{pw} - V_{\alpha p}|$ decreases. We show how the Alfvénic motion of both the alphas and the protons through switchbacks can be understood as approximately rigid arm rotation about the location of the wave frame, and illustrate that the wave frame can therefore be estimated using particle measurements alone, via sphere fitting.
\end{abstract}

\section{Introduction}
One of the more striking results of Parker Solar Probe's (PSP, \cite{fox2016solar}) mission thus far is the ubiquity, in the near-Sun solar wind, of magnetic switchbacks - large, sudden rotations of the magnetic field, accompanied by spikes in the radial solar wind velocity. While switchbacks have previously been observed both in the inner heliosphere using Helios measurements \citep{borovsky2016plasma,horbury2018short}, and at 1AU and beyond \citep{kahler1996topology,neugebauer2013double}, these recent PSP observations have sparked renewed interest in their nature and origins. 

\subsection{Properties}
Switchbacks (hereafter SBs) are long, thin \citep{laker2021statistical,horbury2020sharp}, S-shaped \citep{mcmanus2020cross} magnetic structures, most likely oriented along the magnetic field direction \citep{laker2021statistical}. They are mostly Alfvénic in nature, with constant magnitude $|\mathbf{B}|$ field corresponding to the condition of spherical polarisation. The Alfvénic correlations between $\mathbf{B}$ and $\mathbf{v}$ mean that the field rotations of SBs are accompanied by large positive spikes in the proton velocity, regardless of the underlying polarity of the magnetic field \citep{matteini2014dependence}.
They don't occur continuously but rather appear in ``patches" \citep{bale2019highly,de2020switchbacks}, separated by periods of quiet, steady flow and radial magnetic field. The proton core temperature appears unchanged within individual SBs \citep{woolley2020proton,martinovic2021multiscale}, however the patches themselves appear to be overall hotter than the quiet interstitial periods \citep{woodham2021enhanced,bale2021solar}.

\subsection{SB Formation Theories}
The question of what mechanisms are responsible for switchback generation is still an open one. Several ideas have been put forward, generally coming in two main flavours; the first involves generation via magnetic reconnection. 
\cite{fisk2020global} postulate that due to large scale equatorial circulation of the photospheric magnetic field, open magnetic field lines are dragged across closed loops at lower latitudes, causing interchange reconnection events which launch S-shaped kinks into the corona. \cite{zank2020origin} describe a similar idea but with the reconnection occurring significantly higher up in the corona and launching fast magnetosonic type modes both up and down the open field lines. 

An alternative idea is that switchbacks naturally form in the solar wind as it expands and travels outwards. 
Magnetic field fluctuations decay more slowly with radial distance $R$ than the mean magnetic field does, resulting in normalised amplitudes of Alfvénic fluctuations increasing as a function of $R$. This means that out of the bath of initially small amplitude, linear Alfvén waves known to be present at the base of the corona, the normalised fluctuation amplitudes grow as the plasma travels outwards until they eventually become large enough to cause the field to switch back on itself. \cite{mallet2021evolution} develop an analytical model of such large-amplitude Alfvén waves in an expanding solar wind and make several testable predictions for the properties of the SBs produced, and similar results have been found via MHD simulations \citep{squire2020situ,shoda2021turbulent}.
In-situ generation of SBs then very naturally explains the observation that the SB filling fraction increases as a function of radius \citep{badman2020measurement,mozer2020switchbacks,macneil2020evolution}, something that is difficult to explain for theories involving a purely low coronal origin. It would also explain switchback ``patches" as corresponding to wind that has undergone greater expansion in transit; \cite{bale2021solar} provide strong evidence that at least some of the patches observed by PSP so far are due to superradially expanded wind originating from the boundaries of supergranules at the solar surface. 

Recent analysis of Ulysses, Helios and PSP data by \cite{tenerani2021evolution}  however suggests that the scaling of SB occurrence as a function of radial distance $R$ in fact depends on the size or duration of the switchback, with shorter duration SBs decaying with $R$ and longer ones persisting. This, along with the non-uniform properties of SBs (wide range of durations \citep{de2020switchbacks}, some exhibiting compressibility while most do not \citep{krasnoselskikh2020localized}, different types of discontinuity at the boundaries \citep{larosa2020switchbacks}, etc) could be evidence that both types of generation mechanism are occurring, and we are seeing a combination of short-duration SBs naturally decaying via processes like parametric decay within a few tens of solar radii \citep{tenerani2020magnetic}, while in-situ generation is replenishing the population of longer duration SBs. At this stage this is still speculative and there are many open questions regarding the formation, evolution and eventual decay of SBs. 

For completeness we mention that there are other potential SB generation mechanisms unrelated to the two just described. \cite{ruffolo2020shear} suggest they may be associated with the onset of shear-driven turbulence at or above the Alfvén critical surface. Velocity shears between adjacent flux tubes can then potentially be large enough to trigger the onset of Kelvin-Helmholtz type instabilities and their associated vorticity roll-ups, producing the large deflections in $\mathbf{B}$ that we observe as SBs. \cite{schwadron2021switchbacks} also postulate that SBs are produced by shear interactions between fast and slow streams above the Alfvén surface (when ram pressure becomes dominant), in particular pointing out that this should occur in the super-Parker spiral type magnetic fields produced by footpoint motion across the leading edges of coronal holes.

In this paper we focus on one small piece of the picture, namely the behaviour of alpha particles inside vs outside individual switchbacks, and whether or not this can help distinguish between any potential generation mechanisms.

\section{Methods}
\subsection{Data}
For this study we focus on PSP's third and fourth encounters (E3 and E4) from Aug 27th to Sep 8th, and 2019 and Jan 23rd to Feb 3rd, 2020, respectively. We use data from the FIELDS magnetometers \citep{bale2016fields} for high resolution magnetic field $\mathbf{B}$ measurements and down-sample to match particle measurement cadences as needed. 3D ion velocity distribution function (VDF) measurements are taken from the SPAN-Ion electrostatic analyser \citep{livi2021span,kasper2016solar}, with proton and alpha counts spectra produced at cadences of 7s and 14s respectively. To the proton channel spectra we fit a bi-Maxwellian to both the core and beam populations, with the proton beam constrained to lie along the magnetic field relative to the core velocity. The alpha channel contains a small ($2\%$) contamination from the proton channel, which manifests as scaled down proton core and beam VDFs in the alpha channel. This was accounted for by taking the previously fitted proton parameters and reducing the density down to fit the extraneous protons. An additional single bi-Maxwellian was then fit to the alpha part of the spectrum and the core alpha particle VDF parameters extracted. The $\sim 2\%$ scaling factor is a free parameter in the fit; it was checked that there was no energy or angle dependence in the contaminant protons so that an overall scaling was sufficient. The uncertainties on the fitted alpha densities are approximately $10\%$.

We will on occasion require proton density measurements. For this we use quasi-thermal noise (QTN) estimates derived from extraction of the plasma line from FIELDS RFS spectra \citep{romeo2021characterization}, and approximate $n_e = n_p + 2n_\alpha \approx n_p$, as the alpha abundance in PSP's early encounters is very low \citep{woolley2021plasma}.

\subsection{Switchbacks}
\begin{figure*}[ht!]
    \includegraphics[width=\textwidth]{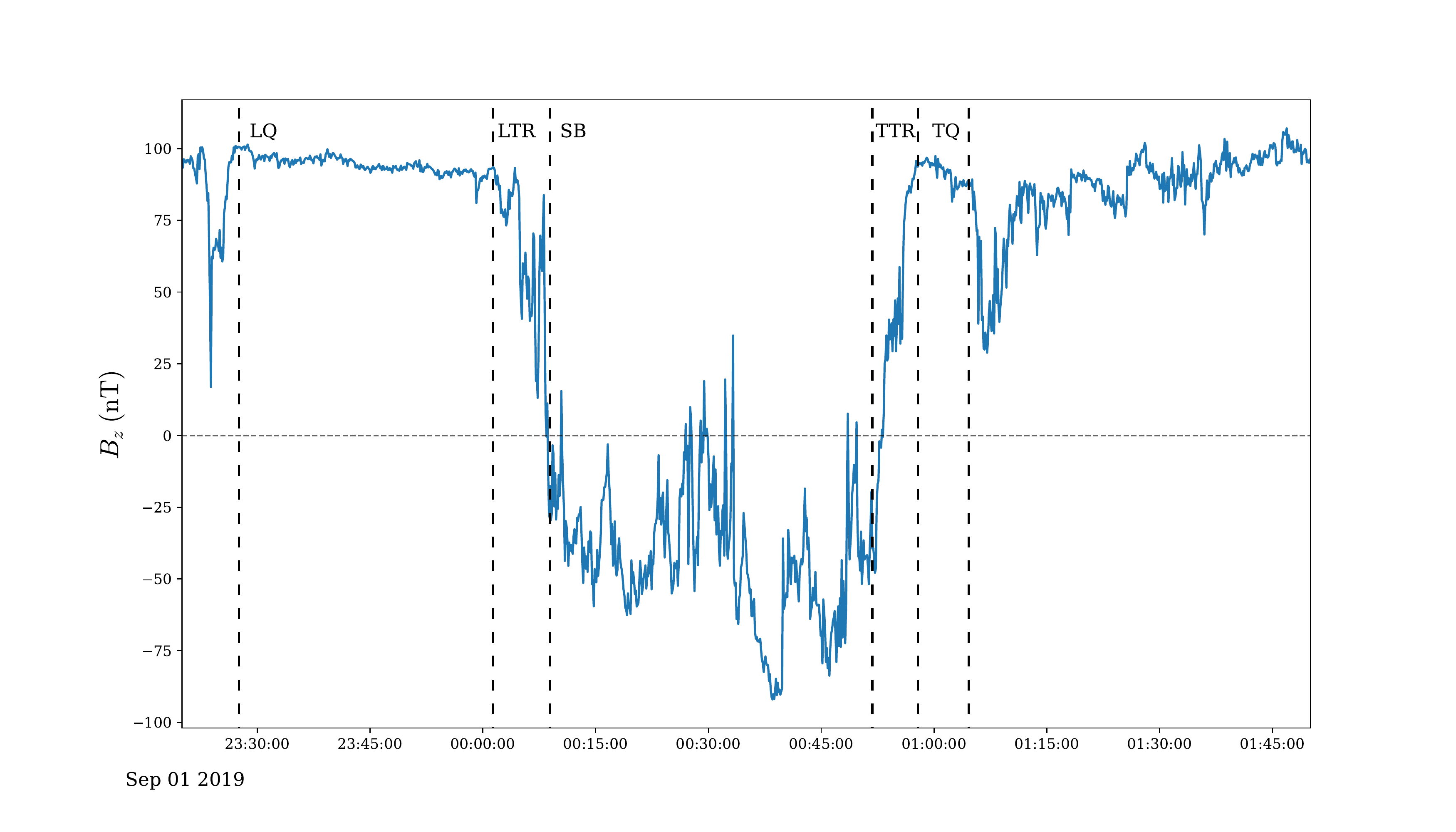}
    \caption{Z-component (in PSP spacecraft coordinates) of the magnetic field  for a typical switchback, showing the demarcation of different regions: Leading Quiet (LQ), Leading Transition Region (LTR), Switchback interior (SB), Trailing Transition Region (TTR), and Trailing Quiet (TQ).}
    \label{fig:sb_example}
\end{figure*}

Our dataset of SBs consists of 92 examples chosen by visual inspection from E3 and E4. Using SPAN-Ion data as the source of our ion measurements means we are constrained by the alpha particle 14s measurement cadence to selecting relatively longer SBs, and so cannot use quite as large an event database as in some previous studies \citep{martinovic2021multiscale}. 
Following \cite{martinovic2021multiscale} we split each SB into five distinct regions: Leading Quiet (LQ),  the relatively quiescent period immediately preceding the switchback; Leading Transition (LT), the transition corresponding to the rotation of the magnetic field; Switchback (SB), the interior of the switchback structure; Trailing Transition (TT), the second transition, and finally the Trailing Quiet (TQ) region, the quiescent field immediately following the passing of the SB. In this work we are mainly interested in comparing the quiescent ``background" conditions to the interior of the switchback, rather than the transition regions which represent the edges of the magnetic structure (and display a host of interesting physics, including signatures of reconnection \citep{froment2021direct} and wave activity \citep{agapitov2020sunward,krasnoselskikh2020localized}). Figure \ref{fig:sb_example} shows a prototypical example SB, with the five regions indicated with vertical dashed lines. 

\section{Results and Discussion}
\subsection{Density and Abundance Changes}

\begin{figure*}
\includegraphics[height=3in]{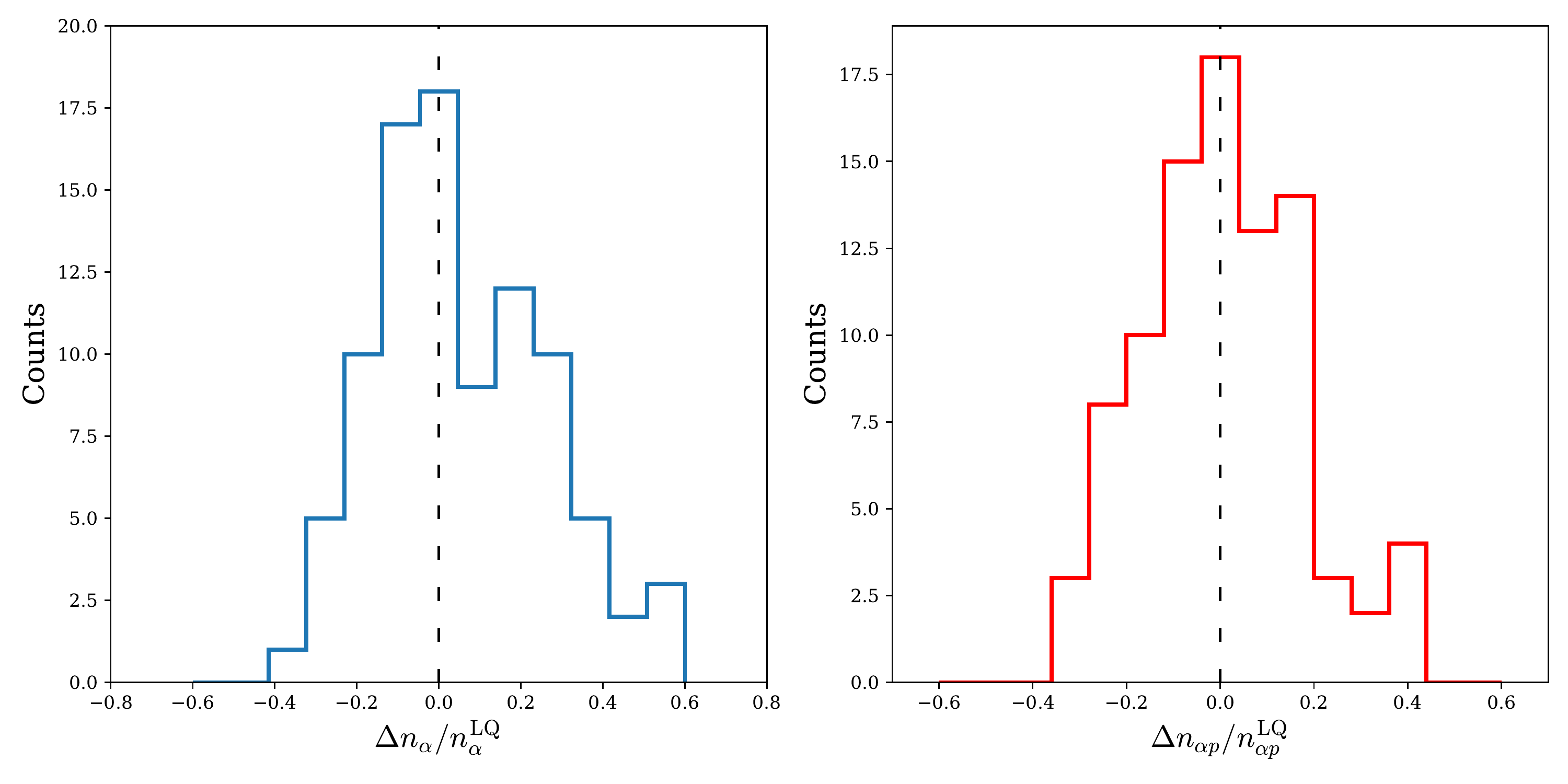}
\centering
\caption{Histograms of fractional change in alpha density (left) and alpha abundance (right) between switchback interiors and their leading quiet regions.}
\label{fig:density_hists}
\end{figure*}

The left plot in figure \ref{fig:density_hists} shows a histogram of the fractional change in alpha number density between SB interiors and their leading quiet (LQ) regions, $(n_\alpha^{\mathrm{SB}} - n_\alpha^{\mathrm{LQ}}) / n_\alpha^{\mathrm{LQ}}$. While the spread in fractional density changes is quite large, the mean (and median) of $\Delta n_\alpha / n_\alpha^{\mathrm{LQ}}$ are both very close to zero (0.05 and 0.02 respectively). This is qualitatively very similar to the proton fractional density changes in SBs reported both in observations and simulations (see Figure 4  of \cite{larosa2020switchbacks} and Figure 10 of \cite{shoda2021turbulent} respectively).
%
%

The histogram on the right in figure \ref{fig:density_hists} shows the change in alpha abundance $\Delta n_{\alpha p} = n_{\alpha}^{\mathrm{SB}}/n_{p}^{\mathrm{SB}} - n_{\alpha}^{\mathrm{LQ}}/n_{p}^{\mathrm{LQ}}$ between the same two regions (note that although one might have $\langle \Delta n_p\rangle \approx 0$ and $\langle \Delta n_\alpha \rangle \approx 0$, a priori they need not be statistically independent). For proton densities we do not use SPAN-Ion measurements of $n_p$, but rather estimates of $n_p$ from FIELDS QTN measurements as detailed in section 2. The large $\delta \mathbf{V}$ associated with SBs often moves the proton VDF significantly out of SPAN-Ion's field of view (FOV), which results in a large (unphysical) proton density decrease as measured by SPAN.
While fitting does mitigate the problem somewhat, it is often not enough to completely eliminate these instrumental density decreases. Using SPAN-Ion measurements only would then appear to show large spikes in the alpha abundance inside SBs compared to outside (not plotted here), which as we have shown is not the case. As we will explain in the next section, the alpha particle VDFs tend to move much less in velocity space during SBs, and so the problem is much less significant and any motion that does occur can be properly captured by the fitting routines. Again, while the spread in the right histogram of figure 2 is relatively large, the distribution is clearly peaked about $\Delta n_{\alpha p}/n_{\alpha p} \approx 0$. We interpret these two figures then as showing there is no statistically significant change in either the alpha density or the alpha abundance inside SBs vs outside. 
\begin{figure}
    \includegraphics[width=0.5\textwidth]{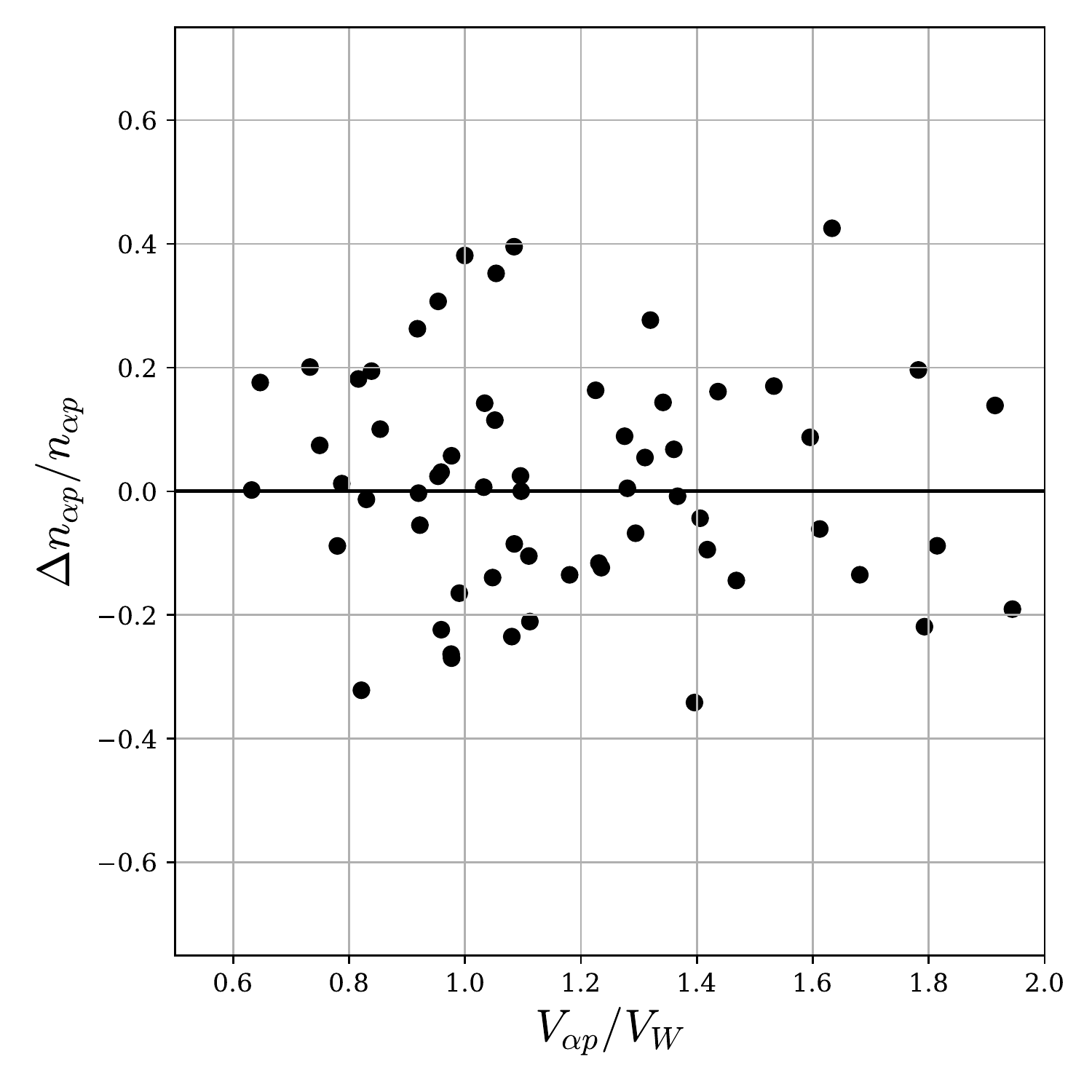}    
    \caption{Scatter plot of fractional change in alpha abundance in switchbacks vs alpha proton drift as a fraction of the local Alfvén wave phase speed. The distribution is symmetric showing no strong dependence.}
    \label{fig:abundance_vs_speed}
\end{figure}

Lack of a compositional signature difference between the SB and LQ regions strongly suggests we are measuring the same plasma inside vs outside, in agreement with previous interpretations of SBs \citep{yamauchi2004differential,mcmanus2020cross,martinovic2021multiscale,woolley2020proton}. We would certainly expect SBs generated in-situ to not display any compositional differences in the plasma inside the SB compared to outside. However, these observations do not rule out coronal origins of SBs. SBs generated by interchange reconnection events further down in the corona may very well be expected to display compositional differences at the time they are generated. This is because the properties of plasma confined in closed magnetic loops is known to change (relative to open field lines) over the confinement time, due to processes like gravitational settling and the first ionization potential (FIP) effect \citep{laming2019element,rakowski2012origin}. However the only way this would be measurable at PSP is if the alpha particles and the SB travel outwards together at exactly the same speed from their point of origin, preserving the compositional signature difference. While it has long been generally understood that alpha particles \emph{do} travel faster than the protons at approximately the local wave speed \citep{thieme1990spatial,steinberg1996differential,matteini2015ion}, giving rise to the phenomenon of alpha particle ``surfing" whereby alpha particles are less affected by the Alfvénic fluctuations, we now show that this isn't always the case, and that expecting a compositional signature to persist to PSP distances would require rather unphysical fine-tuning.

In fig. \ref{fig:abundance_vs_speed} we plot the change in alpha abundance $\Delta n_{\alpha p}$ vs the ratio of alpha-proton drift speed to wave speed, $V_{\alpha p}/V_W$. $V_{\alpha p}$ is calculated as $|V_\alpha - V_{pc}|$ where $V_\alpha$ and $V_{pc}$ are the alpha and proton core velocities respectively, and $V_W$ is computed by taking the normal N-component of the equation
\begin{equation}
    \delta \mathbf{V} = \pm V_W \frac{\delta \mathbf{B}}{|\mathbf{B}|},
    \label{eq:wave_speed}
\end{equation}
which serves to define the wave speed \citep{goldstein1995alfven}.  Plotting $V_N$ vs $B_N/|\mathbf{B}|$ over the LQ region associated with each SB and taking the gradient of a line of best fit then yields an estimate of the local wave phase speed. (Note that equation \ref{eq:wave_speed} is effectively an empirical measurement of the speed of Alfvénic fluctuations - it is not yet fully understood why $V_W$ is usually less than $V_A$ in the solar wind \citep{goldstein1995alfven,neugebauer1996ulysses}.)

From figure \ref{fig:abundance_vs_speed} we clearly see that alpha particles do not always travel at the local wave speed; when considering short intervals such as these, there is a very wide range of $V_{\alpha p}/V_W$ values. There also does not appear to be any trend in $\Delta n_{\alpha p}$ with $V_{\alpha p}/V_W$. In particular, there is no signature around $V_{\alpha p}/V_W \approx 1$, where one might expect such a compositional signature to be were it present when the SB was generated; the spread in points around $V_{\alpha p}/V_W \approx 1$ appears no different than the spread at other values. In retrospect however this is not too surprising, for two reasons. First, even in the model of \cite{zank2020origin} where the interchange reconnection is occurring relatively high up (compared to the photospheric reconnection models of \cite{fisk2020global} and \cite{drake2021switchbacks}), in coronal loops with scale height $\sim 6 R_\odot$, the local Alfvén speed is very high ($V_A \gtrsim 1000$ km/s), and the alpha particles are not expected to ever drift at such high speeds ahead of the protons. Rather, the phenomenon of alpha particles surfing at the Alfvén speed is only expected to kick in at greater radial distances, once the Alfvén speed has decayed enough to be comparable to $V_{\alpha p}$ (and after which it may act as an instability threshold preventing $V_{\alpha p} \gg V_A$ \citep{verscharen2013instabilities}). Thus, we wouldn’t expect $V_{\alpha p}/V_W \approx 1$ to be possible at the site of interchange reconnection, and the alphas would not be able to carry a compositional signature with the SB to be observed at PSP. Secondly, even if the alpha particles could leave the interchange reconnection event at the same speed as the SB, a PSP encounter with perihelion distance $\sim 30 R_\odot$ still represents a travel distance of several hundred Alfvén crossing times (using a typical SB length scale $l \sim 5 \times 10^4$ km \citep{laker2021statistical,tenerani2020magnetic}). Therefore, barring some rather unphysical fine-tuning, any compositional signature would have long since decayed away by the time the alpha particles reached PSP, and we would expect to observe something like figure \ref{fig:abundance_vs_speed}. In conclusion then, our results are all consistent with in-situ generation mechanisms of SBs, but cannot be used to rule out origin mechanisms occurring further down in the corona or at the surface of the Sun.



\subsection{Alfvénic Motion of the Alphas}
\begin{figure*}
    \includegraphics{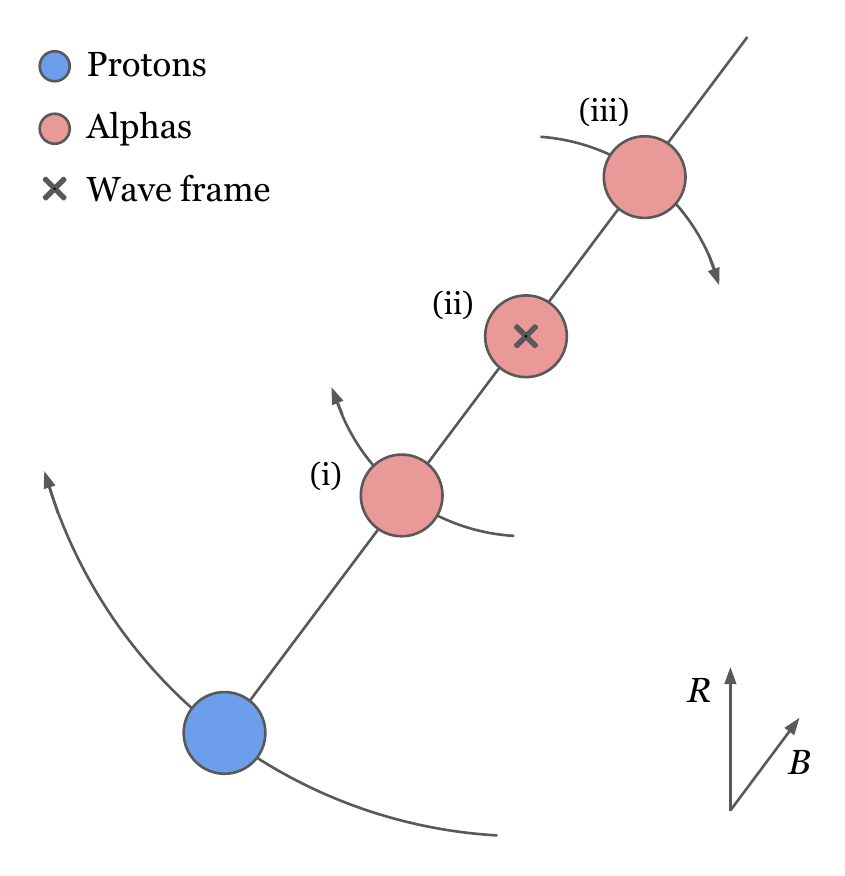}
    \centering
    \caption{Cartoon showing the idealised expected alpha particle motion in velocity space during a SB in the three scenarios (i) $V_{\alpha p} < V_w$, (ii) $V_{\alpha p} \sim V_w$, (iii) $V_{\alpha p} > V_w$, corresponding to the rows of figure \ref{fig:sb_spheres}. $R$ and $B$ denote the radial and magnetic field directions respectively.}
    \label{fig:cartoon}
\end{figure*}
SBs are known to be highly Alfvénic and spherically polarised ($|\mathbf{B}| = \text{constant}$), and we therefore expect the particle motion to be spherically polarised too. To see why, consider a particle at rest in the frame co-moving with the Alfvén wave. The magnetic field, being Galilean invariant, is still spherically polarised, and the wave being stationary means that energy is conserved in this frame (and that the electric field should almost vanish). A particle with perturbed velocity $\delta \mathbf{v}$ relative to this frame must therefore trace out a sphere in velocity space in order to conserve energy. Boosting back into the spacecraft frame we infer that the observed motion should be spherically polarised, centred at the wave frame velocity, with radius equal to the wave speed relative to whichever particle population we are considering. (For a more in-depth discussion of this see \cite{matteini2015ion}.) With this picture in mind, one can potentially expect three different types of alpha particle motion, depending on the relative magnitudes of $V_{\alpha p}$ and $V_{pw}$; these are sketched out in the cartoon in figure \ref{fig:cartoon}. In scenario (i) we have $V_{\alpha p} < V_{pw}$, and would expect to observe spherical polarisation of both the protons and alpha particles, with the alpha particles tracing out a sphere of smaller radius than the protons, approximately given by $V_{\alpha w} \approx V_{pw} - V_{\alpha p}$. In case (ii), the position of the alphas in velocity space roughly coincides with the wave frame, $V_{pw} \approx V_{\alpha p}$, and one would expect the protons to be spherically polarised but the alphas to be roughly stationary. In case (iii) we have $V_{\alpha p} > V_{pw}$, and so would again expect the protons and alphas to be spherically polarised, but importantly the alphas should move in anti-phase with the protons. This potential for alphas to move either in-phase or in anti-phase with protons during Alfvénic fluctuations depending on the relative values of $V_{\alpha p}$ and $V_{pw}$ was first pointed out by \cite{goldstein1995alfven} using Ulysses data. Understanding this in terms of spherical motion of each species in velocity space is exactly the model laid out in \cite{matteini2015ion}, the only difference here is that the cadence and quality of the SPAN-Ion measurements allow us to distinguish between the three cases over short timescales, and directly observe and measure the spherical polarisation of the alphas. 
\begin{figure*}
    \includegraphics[width=0.8\textwidth]{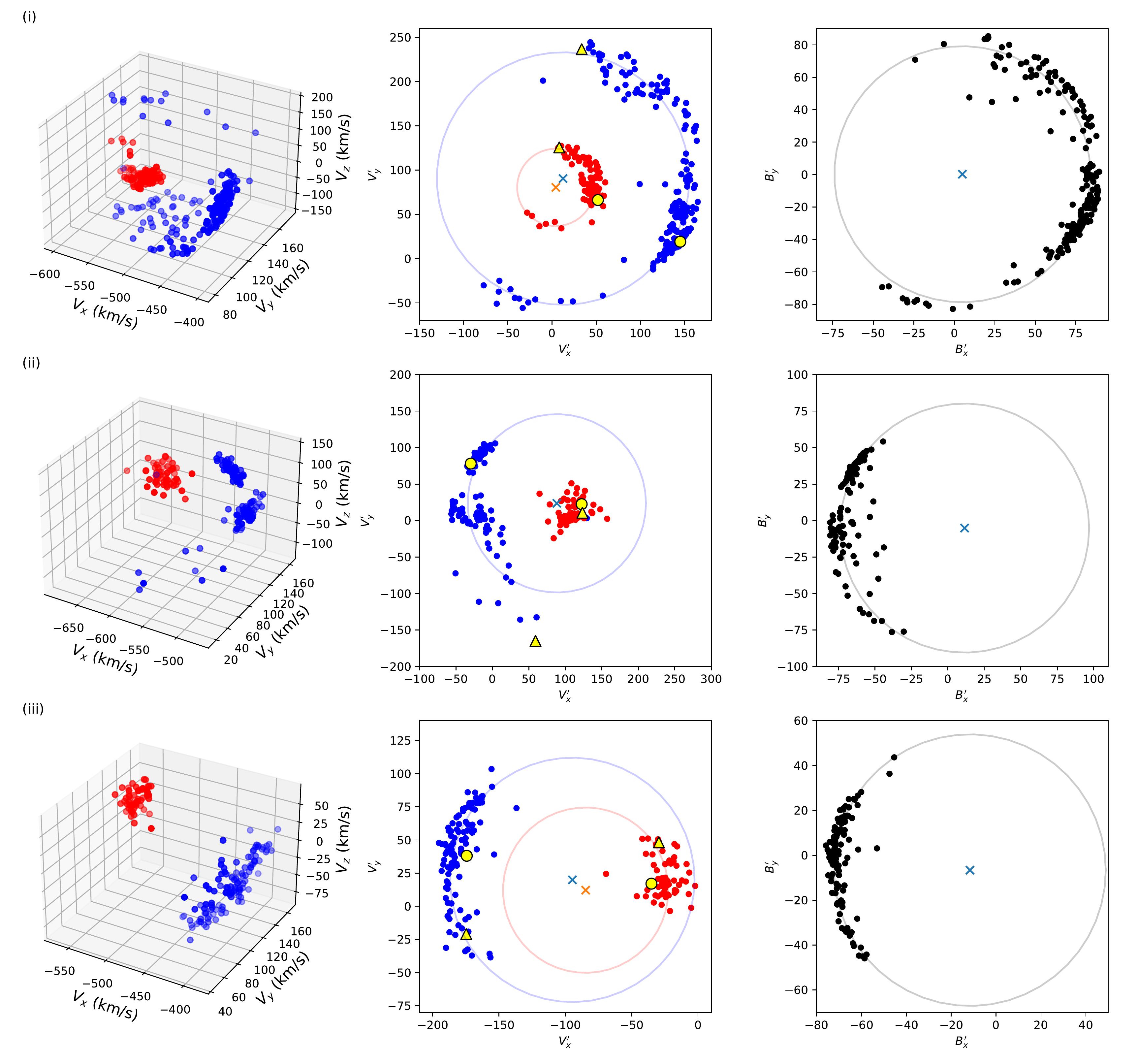}
    \centering
    \caption{Three example SBs showing the different types of alpha-particle Alfvénic motion. First column is the 3D proton and alpha velocity measurements in instrument coordinates, in blue and red respectively, through the switchback. Middle column are these particle velocities projected onto the minimum variance magnetic field plane. Yellow circles indicate the start of the SB interval, yellow triangles the point of maximum deflection during the SB. Third column is the magnetic field measurements projected onto the same plane, with circles of best fit in grey (blue crosses mark the circles' centres).}
    \label{fig:sb_spheres}
\end{figure*}
In each row of figure \ref{fig:sb_spheres} we show an example SB, illustrating the three main types of alpha particle motion just described. The left column shows the 3D proton and alpha velocity measurements (in blue and red respectively), in instrument coordinates. In all three cases the proton motion (in blue) is spherically polarised as expected. Regarded as single large-amplitude, low-frequency Alfvén waves, these switchbacks are not just spherically polarised but to a good approximation arc-polarised as well \citep{fisk2020global}, as first theoretically predicted by \cite{barnes1974large} and observed many times in the solar wind since \citep{lichtenstein1980dynamic,tsurutani1994relationship,riley1996properties}. Their maximum and intermediate principal component axes define a plane that is almost constant through the SB interval, and the tip of the $\mathbf{B}$ field roughly traces out an arc on the sphere of $|\mathbf{B}| = \text{constant}$. If $\mathbf{e}_1,\mathbf{e}_2,\mathbf{e}_3$ are the orthonormal principal components of the magnetic field measurements $\mathbf{B}_i$ for a single SB interval with eigenvalues $\lambda_1 \lesssim \lambda_2 \ll \lambda_3$, we can project the measurements onto the plane defined by $e_2$ and $e_3$, which should be the plane in which they appear most circular; this is shown in the third column of fig. \ref{fig:sb_spheres}. The fluctuations being Alfvénic means we can project the velocity measurements onto the same plane; this is plotted in the second column of fig. \ref{fig:sb_spheres}. 

From the middle plot of the first SB example (taken from 2019-08-30/22:50:25 to 2019-08-30/23:12:47), we can see that the alpha particle velocities in red are spherically polarised as well, albeit with a smaller amplitude - they move on the surface of a smaller sphere. To a good approximation, the alpha and proton velocities both subtend the same angle $\theta$ that the magnetic field does, and appear to be rotating about a similar point in velocity space. The yellow circle represents the start of the entire SB interval, and the yellow triangle the point of maximum $\mathbf{B}$ field deflection during the SB. From these we can see that the protons and alphas are moving in phase with each other. This corresponds to scenario (i), $V_{\alpha p} < V_{pw}$, in figure \ref{fig:cartoon}.
Circles of best fit to the proton and alpha motion are overlaid in blue and red respectively, their centres marked with crosses. The proximity of the centres of the alpha and proton circles shows good agreement between these two components of the wave frame velocity. The third component can be estimated by calculating the $(e_2,e_3)$ plane that minimises the least square distance to the measured velocities for each species separately (this is then the plane of arc-polarisation). For protons the sphere centre is $\mathbf{v}_{pw} = (-543,147,16) \text{ km/s}$ and for the alphas it is $\mathbf{v}_{\alpha w} = (-547,152,28) \text{ km/s}$ (in instrument coordinates), showing very good agreement in all three components. Because the alpha fits are independent of the proton fits (they are not constrained to lie along the magnetic field relative to the proton VDFs), fitting spheres in this way represents two independent estimates of the wave frame velocity. 

Using equation \ref{eq:wave_speed} during the LQ region of this SB, we estimate the local alpha and proton wave phase speeds as $V_{pw} \approx 149$ km/s and $V_{\alpha w} \approx 45$ km/s. This is in excellent agreement with the radii of the spheres of best fit in column 2, which have radii of 143 and 44 km/s respectively. With $V_{\alpha p} \approx 114$ km/s in the LQ region we also have $V_{\alpha p} < V_{pw}$ and $V_{\alpha p} + V_{\alpha w} \approx V_{pw}$ as expected. Comparing to the measured Alfvén speed $V_A \approx 147$ km/s, for this SB we have that the proton phase speed and Alfvén speed are almost equal, $V_{pw} \approx 1.01 V_A$. In time series of the particle velocities for SBs like this, the large spikes in the proton velocity would also be seen in the alphas, albeit smaller in magnitude.

In the second example SB (middle row), (taken from 2019-08-29/21:08:04 to 2019-08-29/21:20:46), the proton velocities are still spherically polarised, but the alpha’s are not - they appear relatively stationary in velocity space through the SB and do not trace out an arc (the yellow markers for initial and maximum $\mathbf{B}$ deflection lie almost on top of each other).
This corresponds to scenario (ii) in figure \ref{fig:cartoon}, where $V_{\alpha w} \approx 0$, and for the LQ interval preceding this SB, we have $V_{\alpha p} \approx 158$ km/s, $V_{pw} \approx 162$ km/s, and $V_{\alpha w} \approx 6$ km/s, with $V_A \approx 161$ km/s so that $V_{pw} \approx 1.01 V_A$. The alphas are therefore roughly comoving with the wave, and their location in velocity space serves as an estimate of the wave frame. The centre of the proton circle of best fit in blue lies reasonably close to the alpha velocities, but we note there is a fair amount of scatter in the proton measurements for this SB.

Finally, in the third example SB in the bottom row (from 2019-08-29/08:37:52 to 2019-08-29/08:51:09), the alphas are again spherically polarised, but moving in anti-phase with the protons, as can be seen by the relative locations of the points of maximum SB deflection (yellow triangles). This corresponds to scenario (iii) in figure \ref{fig:cartoon}. For this LQ region we have $V_{\alpha p} \approx 167$ km/s, $V_{pw} \approx 101$ km/s, and $V_{\alpha w} \approx -26$ km/s, with $V_A \approx 105$ km/s so that $V_{pw} \approx 0.96 V_A$. The quantitative agreement is not quite as good as previous, qualitatively however $V_{\alpha p} > V_{pw}$, and $V_{\alpha w}$ and $V_{pw}$ have opposite signs, as expected. For SBs such as these, a time series of particle velocities would see spikes in proton velocity coinciding with dips in alpha velocity. We note that this anti-phase motion would also be expected to be observed in proton beams, since they typically travel at or slightly above the Alfvén speed relative to the core \citep{alterman2018comparison}.

\subsection{Relation to the de Hoffman-Teller Frame}
Finally, for completeness we compare the wave frames determined using the methods described above with the direct computation of the de Hoffman-Teller (DHT) frame. First introduced by \cite{de1950magneto} in the context of MHD shocks, it is defined to be the frame in which the plasma's electric field vanishes, and is usually computed by finding the velocity $\mathbf{V}$ that minimizes the quantity \citep{khrabrov1998dehoffmann}
\begin{equation}
    \label{eq:min_DHT}
    D(\mathbf{V}) = \frac{1}{M} \sum_{m=1}^M \left \vert (\mathbf{v}^{(m)} - \mathbf{V})\times \mathbf{B}^{(m)} \right \vert^2,
\end{equation}
where $\mathbf{v}^{(m)}, \mathbf{B}^{(m)}$ denote velocity and magnetic field values over a series of measurements indexed by $m = 1,\dots,M$.
By definition we expect the DHT frame and the wave frames computed above to be one and the same. To see geometrically why this is so for the SBs being considered here, consider the ideal case of a perfectly spherically polarised Alfvén wave. The minimum value of $D(\mathbf{V}) = 0$ will be achieved only if each term in eq. \ref{eq:min_DHT} vanishes, which requires $\mathbf{V}_{\text{DHT}}$ to lie on the line through $\mathbf{v}^{(m)}$ parallel to $\mathbf{B}^{(m)}$, for each measurement $m$. The point that uniquely satisfies this is the centre of the sphere in velocity space (as it is the point of intersection of each of these lines through $\mathbf{v}^{(m)}$). Thus, regarding SBs as essentially single, large-amplitude spherically polarised Alfvén waves, we expect $\mathbf{V}_{\text{DHT}}$ and $\mathbf{v}_{pw}$ to agree to good approximation, and this is indeed the case for our three example SBs, as summarised in Table \ref{tab:wave_speeds}. 


\begin{deluxetable}{c c c}
    \tablecaption{Wave speeds determined by sphere fitting vs $V_{DHT}$}\label{tab:wave_speeds}
    \tablehead{\colhead{SB} & \colhead{$\mathbf{v}_{pw}$} & \colhead{$\mathbf{V}_{DHT}$} \\ \colhead{} & \colhead{(km/s)}&  \colhead{(km/s)}}
    \startdata
    (i) & (-543, 147, 16) & (-548, 116, 9) \\ 
    (ii) & (-598, 155, 33) & (-614, 152, 12) \\
    (iii) & (-483, 128, 1) & (-473, 123, 9) \\
    \enddata
\end{deluxetable}
 

\section{Conclusions}
In this work the density and abundance variations of alpha particles were examined in a database of 92 switchbacks from PSP’s encounter 3 and 4. No consistent compositional signature difference was observed in the alpha abundance $n_{\alpha p}$ inside SBs vs outside, suggesting that PSP is measuring the same plasma in both cases, in agreement with previous interpretations of SBs \citep{yamauchi2004differential,martinovic2021multiscale,woolley2020proton}. We argued that even if SBs are the results of interchange reconnection events lower down in the corona, compositional signatures are not likely to exist and be measurable at PSP for two reasons: 1) the local Alfvén speed at the postulated interchange reconnection sites is very high and most likely precludes alphas being able to travel with the SBs that are launched upwards along the field lines (thus preventing compositional information being carried with the SB), and 2) even if the alphas are able to travel with the SB, a small difference between $V_{\alpha p}$ and $V_{pw}$ would cause a compositional signature to have long decayed away due to the distance (in Alfvén crossing times) to PSP’s perihelia. Thus, our observation of there being no dependence of $\Delta n_{\alpha p}$ on $V_{\alpha p}/V_{pw}$ is to be expected and does not help distinguish between in-situ generation and interchange reconnection as potential SB formation mechanisms.


In addition, we examined the three-dimensional nature of the velocity fluctuations of both protons and alphas within individual SBs. We observed spherical polarisation of both the proton and alpha velocities, which can be understood as a consequence of energy conservation in the wave frame. Three example SBs showed the alphas moving in-phase, stationary relative to, and in anti-phase with, the protons. This corresponds to the three cases $V_{\alpha p} < V_{pw}$, $V_{\alpha p} \approx V_{pw}$, and $V_{\alpha p} > V_{pw}$. Thus while SBs are always associated with spikes in the proton velocity, alpha velocities may be enhanced, unchanged, or decrease, depending on the relative values of $V_{\alpha p}$ and $V_{pw}$. For the case $V_{\alpha p} < V_{pw}$, where the alphas move in phase on a sphere of smaller radius than the protons, the centres of the proton and alpha velocity spheres were in excellent agreement, illustrating how one can make two independent particle measurements to uniquely identify the wave frame. One can in principle use these methods to estimate the wave frame over short time scales using purely particle measurements, and we showed that this agreed well with the the usual method of computing the de Hoffman-Teller frame via minimisation of the motional electric field, $\mathbf{E} = - \mathbf{v} \times \mathbf{B}$.
Intuitively then the Alfvenic motion of both the alphas and the protons through SBs can be understood as approximately rigid arm rotation about the location of the wave frame in velocity space, as illustrated in \ref{fig:cartoon} and discussed in \cite{matteini2014dependence,matteini2015ion}, and the length of the lever arms are to a good approximation given by $V_{pw}$ and $V_{\alpha w}$ for protons and alphas respectively.


\bibliography{biblio}

\begin{thebibliography}{}
\expandafter\ifx\csname natexlab\endcsname\relax\def\natexlab#1{#1}\fi
\providecommand{\url}[1]{\href{#1}{#1}}
\providecommand{\dodoi}[1]{doi:~\href{http://doi.org/#1}{\nolinkurl{#1}}}
\providecommand{\doeprint}[1]{\href{http://ascl.net/#1}{\nolinkurl{http://ascl.net/#1}}}
\providecommand{\doarXiv}[1]{\href{https://arxiv.org/abs/#1}{\nolinkurl{https://arxiv.org/abs/#1}}}

\bibitem[{Agapitov {et~al.}(2020)Agapitov, de~Wit, Mozer, Bonnell, Drake,
  Malaspina, Krasnoselskikh, Bale, Whittlesey, Case,
  {et~al.}}]{agapitov2020sunward}
Agapitov, O., de~Wit, T.~D., Mozer, F., {et~al.} 2020, The Astrophysical
  journal letters, 891, L20

\bibitem[{Alterman {et~al.}(2018)Alterman, Kasper, Stevens, \&
  Koval}]{alterman2018comparison}
Alterman, B., Kasper, J.~C., Stevens, M.~L., \& Koval, A. 2018, The
  Astrophysical Journal, 864, 112

\bibitem[{{Badman, Samuel T.} {et~al.}(2021){Badman, Samuel T.}, {Bale, Stuart
  D.}, {Rouillard, Alexis P.}, {Bowen, Trevor A.}, {Bonnell, John W.}, {Goetz,
  Keith}, {Harvey, Peter R.}, {MacDowall, Robert J.}, {Malaspina, David M.}, \&
  {Pulupa, Marc}}]{badman2020measurement}
{Badman, Samuel T.}, {Bale, Stuart D.}, {Rouillard, Alexis P.}, {et~al.} 2021,
  A\&A, 650, A18, \dodoi{10.1051/0004-6361/202039407}

\bibitem[{Bale {et~al.}(2016)Bale, Goetz, Harvey, Turin, Bonnell, De~Wit,
  Ergun, MacDowall, Pulupa, Andr{\'e}, {et~al.}}]{bale2016fields}
Bale, S., Goetz, K., Harvey, P., {et~al.} 2016, Space science reviews, 204, 49

\bibitem[{Bale {et~al.}(2019)Bale, Badman, Bonnell, Bowen, Burgess, Case,
  Cattell, Chandran, Chaston, Chen, {et~al.}}]{bale2019highly}
Bale, S., Badman, S., Bonnell, J., {et~al.} 2019, Nature, 576, 237

\bibitem[{Bale {et~al.}(2021)Bale, Horbury, Velli, Desai, Halekas, McManus,
  Panasenco, Badman, Bowen, Chandran, {et~al.}}]{bale2021solar}
Bale, S., Horbury, T., Velli, M., {et~al.} 2021, The Astrophysical Journal,
  923, 174

\bibitem[{Barnes \& Hollweg(1974)}]{barnes1974large}
Barnes, A., \& Hollweg, J.~V. 1974, Journal of Geophysical Research, 79, 2302

\bibitem[{Borovsky(2016)}]{borovsky2016plasma}
Borovsky, J.~E. 2016, Journal of Geophysical Research: Space Physics, 121, 5055

\bibitem[{De~Hoffmann \& Teller(1950)}]{de1950magneto}
De~Hoffmann, F., \& Teller, E. 1950, Physical Review, 80, 692

\bibitem[{de~Wit {et~al.}(2020)de~Wit, Krasnoselskikh, Bale, Bonnell, Bowen,
  Chen, Froment, Goetz, Harvey, Jagarlamudi, {et~al.}}]{de2020switchbacks}
de~Wit, T.~D., Krasnoselskikh, V.~V., Bale, S.~D., {et~al.} 2020, The
  Astrophysical Journal Supplement Series, 246, 39

\bibitem[{Drake {et~al.}(2021)Drake, Agapitov, Swisdak, Badman, Bale, Horbury,
  Kasper, MacDowall, Mozer, Phan, {et~al.}}]{drake2021switchbacks}
Drake, J., Agapitov, O., Swisdak, M., {et~al.} 2021, Astronomy \& Astrophysics,
  650, A2

\bibitem[{Fisk \& Kasper(2020)}]{fisk2020global}
Fisk, L., \& Kasper, J. 2020, The Astrophysical Journal Letters, 894, L4

\bibitem[{Fox {et~al.}(2016)Fox, Velli, Bale, Decker, Driesman, Howard, Kasper,
  Kinnison, Kusterer, Lario, {et~al.}}]{fox2016solar}
Fox, N., Velli, M., Bale, S., {et~al.} 2016, Space Science Reviews, 204, 7

\bibitem[{Froment {et~al.}(2021)Froment, Krasnoselskikh, de~Wit, Agapitov,
  Fargette, Lavraud, Larosa, Kretzschmar, Jagarlamudi, Velli,
  {et~al.}}]{froment2021direct}
Froment, C., Krasnoselskikh, V., de~Wit, T.~D., {et~al.} 2021, Astronomy \&
  Astrophysics, 650, A5

\bibitem[{Goldstein {et~al.}(1995)Goldstein, Neugebauer, \&
  Smith}]{goldstein1995alfven}
Goldstein, B., Neugebauer, M., \& Smith, E. 1995, Geophysical research letters,
  22, 3389

\bibitem[{Horbury {et~al.}(2018)Horbury, Matteini, \&
  Stansby}]{horbury2018short}
Horbury, T., Matteini, L., \& Stansby, D. 2018, Monthly Notices of the Royal
  Astronomical Society, 478, 1980

\bibitem[{Horbury {et~al.}(2020)Horbury, Woolley, Laker, Matteini, Eastwood,
  Bale, Velli, Chandran, Phan, Raouafi, {et~al.}}]{horbury2020sharp}
Horbury, T.~S., Woolley, T., Laker, R., {et~al.} 2020, The Astrophysical
  Journal Supplement Series, 246, 45

\bibitem[{Kahler {et~al.}(1996)Kahler, Crocker, \&
  Gosling}]{kahler1996topology}
Kahler, S., Crocker, N., \& Gosling, J. 1996, Journal of Geophysical Research:
  Space Physics, 101, 24373

\bibitem[{Kasper {et~al.}(2016)Kasper, Abiad, Austin, Balat-Pichelin, Bale,
  Belcher, Berg, Bergner, Berthomier, Bookbinder, {et~al.}}]{kasper2016solar}
Kasper, J.~C., Abiad, R., Austin, G., {et~al.} 2016, Space Science Reviews,
  204, 131

\bibitem[{Khrabrov \& Sonnerup(1998)}]{khrabrov1998dehoffmann}
Khrabrov, A.~V., \& Sonnerup, B.~U. 1998, Analysis methods for multi-spacecraft
  data, 1, 221

\bibitem[{Krasnoselskikh {et~al.}(2020)Krasnoselskikh, Larosa, Agapitov,
  De~Wit, Moncuquet, Mozer, Stevens, Bale, Bonnell, Froment,
  {et~al.}}]{krasnoselskikh2020localized}
Krasnoselskikh, V., Larosa, A., Agapitov, O., {et~al.} 2020, The Astrophysical
  Journal, 893, 93

\bibitem[{Laker {et~al.}(2021)Laker, Horbury, Bale, Matteini, Woolley, Woodham,
  Badman, Pulupa, Kasper, Stevens, {et~al.}}]{laker2021statistical}
Laker, R., Horbury, T.~S., Bale, S.~D., {et~al.} 2021, Astronomy \&
  Astrophysics, 650, A1

\bibitem[{Laming {et~al.}(2019)Laming, Vourlidas, Korendyke, Chua, Cranmer, Ko,
  Kuroda, Provornikova, Raymond, Raouafi, {et~al.}}]{laming2019element}
Laming, J.~M., Vourlidas, A., Korendyke, C., {et~al.} 2019, The Astrophysical
  Journal, 879, 124

\bibitem[{{Larosa, A.} {et~al.}(2021){Larosa, A.}, {Krasnoselskikh, V.}, {Dudok
  de Wit, T.}, {Agapitov, O.}, {Froment, C.}, {Jagarlamudi, V. K.}, {Velli,
  M.}, {Bale, S. D.}, {Case, A. W.}, {Goetz, K.}, {Harvey, P.}, {Kasper, J.
  C.}, {Korreck, K. E.}, {Larson, D. E.}, {MacDowall, R. J.}, {Malaspina, D.},
  {Pulupa, M.}, {Revillet, C.}, \& {Stevens, M. L.}}]{larosa2020switchbacks}
{Larosa, A.}, {Krasnoselskikh, V.}, {Dudok de Wit, T.}, {et~al.} 2021, A\&A,
  650, A3, \dodoi{10.1051/0004-6361/202039442}

\bibitem[{Lichtenstein \& Sonett(1980)}]{lichtenstein1980dynamic}
Lichtenstein, B., \& Sonett, C. 1980, Geophysical Research Letters, 7, 189

\bibitem[{Livi {et~al.}(2021)Livi, Larson, Kasper, Abiad, Case, Klein, Curtis,
  Dalton, Stevens, Korreck, \& et~al.}]{livi2021span}
Livi, R., Larson, D.~E., Kasper, J.~C., {et~al.} 2021, Earth and Space Science
  Open Archive, 20, \dodoi{10.1002/essoar.10508651.1}

\bibitem[{Macneil {et~al.}(2020)Macneil, Owens, Wicks, Lockwood, Bentley, \&
  Lang}]{macneil2020evolution}
Macneil, A.~R., Owens, M.~J., Wicks, R.~T., {et~al.} 2020, Monthly Notices of
  the Royal Astronomical Society, 494, 3642

\bibitem[{Mallet {et~al.}(2021)Mallet, Squire, Chandran, Bowen, \&
  Bale}]{mallet2021evolution}
Mallet, A., Squire, J., Chandran, B.~D., Bowen, T., \& Bale, S.~D. 2021, The
  Astrophysical Journal, 918, 62

\bibitem[{Martinovi{\'c} {et~al.}(2021)Martinovi{\'c}, Klein, Huang, Chandran,
  Kasper, Lichko, Bowen, Chen, Matteini, Stevens,
  {et~al.}}]{martinovic2021multiscale}
Martinovi{\'c}, M.~M., Klein, K.~G., Huang, J., {et~al.} 2021, The
  Astrophysical Journal, 912, 28

\bibitem[{Matteini {et~al.}(2015)Matteini, Horbury, Pantellini, Velli, \&
  Schwartz}]{matteini2015ion}
Matteini, L., Horbury, T., Pantellini, F., Velli, M., \& Schwartz, S. 2015, The
  Astrophysical Journal, 802, 11

\bibitem[{Matteini {et~al.}(2014)Matteini, Horbury, Neugebauer, \&
  Goldstein}]{matteini2014dependence}
Matteini, L., Horbury, T.~S., Neugebauer, M., \& Goldstein, B.~E. 2014,
  Geophysical Research Letters, 41, 259

\bibitem[{McManus {et~al.}(2020)McManus, Bowen, Mallet, Chen, Chandran, Bale,
  Larson, de~Wit, Kasper, Stevens, {et~al.}}]{mcmanus2020cross}
McManus, M.~D., Bowen, T.~A., Mallet, A., {et~al.} 2020, The Astrophysical
  Journal Supplement Series, 246, 67

\bibitem[{Mozer {et~al.}(2020)Mozer, Agapitov, Bale, Bonnell, Case, Chaston,
  Curtis, De~Wit, Goetz, Goodrich, {et~al.}}]{mozer2020switchbacks}
Mozer, F., Agapitov, O., Bale, S., {et~al.} 2020, The Astrophysical Journal
  Supplement Series, 246, 68

\bibitem[{Neugebauer {et~al.}(1996)Neugebauer, Goldstein, Smith, \&
  Feldman}]{neugebauer1996ulysses}
Neugebauer, M., Goldstein, B., Smith, E., \& Feldman, W. 1996, Journal of
  Geophysical Research: Space Physics, 101, 17047

\bibitem[{Neugebauer \& Goldstein(2013)}]{neugebauer2013double}
Neugebauer, M., \& Goldstein, B.~E. 2013, in AIP Conference Proceedings, Vol.
  1539, American Institute of Physics, 46--49

\bibitem[{Rakowski \& Laming(2012)}]{rakowski2012origin}
Rakowski, C.~E., \& Laming, J.~M. 2012, The Astrophysical Journal, 754, 65

\bibitem[{Riley {et~al.}(1996)Riley, Sonett, Tsurutani, Balogh, Forsyth, \&
  Hoogeveen}]{riley1996properties}
Riley, P., Sonett, C., Tsurutani, B., {et~al.} 1996, Journal of Geophysical
  Research: Space Physics, 101, 19987

\bibitem[{Romeo {et~al.}(2021)Romeo, Larson, Whittlesey, Rahmati, Livi, \&
  Badman}]{romeo2021characterization}
Romeo, O., Larson, D.~E., Whittlesey, P.~L., {et~al.} 2021, in AGU Fall Meeting
  2021, AGU

\bibitem[{Ruffolo {et~al.}(2020)Ruffolo, Matthaeus, Chhiber, Usmanov, Yang,
  Bandyopadhyay, Parashar, Goldstein, DeForest, Wan,
  {et~al.}}]{ruffolo2020shear}
Ruffolo, D., Matthaeus, W.~H., Chhiber, R., {et~al.} 2020, The Astrophysical
  Journal, 902, 94

\bibitem[{Schwadron \& McComas(2021)}]{schwadron2021switchbacks}
Schwadron, N., \& McComas, D. 2021, The Astrophysical Journal, 909, 95

\bibitem[{Shoda {et~al.}(2021)Shoda, Chandran, \& Cranmer}]{shoda2021turbulent}
Shoda, M., Chandran, B.~D., \& Cranmer, S.~R. 2021, The Astrophysical Journal,
  915, 52

\bibitem[{Squire {et~al.}(2020)Squire, Chandran, \& Meyrand}]{squire2020situ}
Squire, J., Chandran, B.~D., \& Meyrand, R. 2020, The Astrophysical Journal
  Letters, 891, L2

\bibitem[{Steinberg {et~al.}(1996)Steinberg, Lazarus, Ogilvie, Lepping, \&
  Byrnes}]{steinberg1996differential}
Steinberg, J., Lazarus, A., Ogilvie, K., Lepping, R., \& Byrnes, J. 1996,
  Geophysical research letters, 23, 1183

\bibitem[{Tenerani {et~al.}(2021)Tenerani, Sioulas, Matteini, Panasenco, Shi,
  \& Velli}]{tenerani2021evolution}
Tenerani, A., Sioulas, N., Matteini, L., {et~al.} 2021, The Astrophysical
  Journal Letters, 919, L31

\bibitem[{Tenerani {et~al.}(2020)Tenerani, Velli, Matteini, R{\'e}ville, Shi,
  Bale, Kasper, Bonnell, Case, De~Wit, {et~al.}}]{tenerani2020magnetic}
Tenerani, A., Velli, M., Matteini, L., {et~al.} 2020, The Astrophysical Journal
  Supplement Series, 246, 32

\bibitem[{Thieme {et~al.}(1990)Thieme, Marsch, \& Schwenn}]{thieme1990spatial}
Thieme, K., Marsch, E., \& Schwenn, R. 1990, in Annales Geophysicae, Vol.~8,
  713--723

\bibitem[{Tsurutani {et~al.}(1994)Tsurutani, Ho, Smith, Neugebauer, Goldstein,
  Mok, Arballo, Balogh, Southwood, \& Feldman}]{tsurutani1994relationship}
Tsurutani, B., Ho, C., Smith, E., {et~al.} 1994, Geophysical Research Letters,
  21, 2267

\bibitem[{Verscharen {et~al.}(2013)Verscharen, Bourouaine, \&
  Chandran}]{verscharen2013instabilities}
Verscharen, D., Bourouaine, S., \& Chandran, B.~D. 2013, The Astrophysical
  Journal, 773, 163

\bibitem[{Woodham {et~al.}(2021)Woodham, Horbury, Matteini, Woolley, Laker,
  Bale, Nicolaou, Stawarz, Stansby, Hietala, {et~al.}}]{woodham2021enhanced}
Woodham, L., Horbury, T., Matteini, L., {et~al.} 2021, Astronomy \&
  Astrophysics, 650, L1

\bibitem[{Woolley {et~al.}(2020)Woolley, Matteini, Horbury, Bale, Woodham,
  Laker, Alterman, Bonnell, Case, Kasper, {et~al.}}]{woolley2020proton}
Woolley, T., Matteini, L., Horbury, T.~S., {et~al.} 2020, Monthly Notices of
  the Royal Astronomical Society, 498, 5524

\bibitem[{Woolley {et~al.}(2021)Woolley, Matteini, McManus, Ber{\v{c}}i{\v{c}},
  Badman, Woodham, Horbury, Bale, Laker, Stawarz, {et~al.}}]{woolley2021plasma}
Woolley, T., Matteini, L., McManus, M.~D., {et~al.} 2021, Monthly Notices of
  the Royal Astronomical Society, 508, 236

\bibitem[{Yamauchi {et~al.}(2004)Yamauchi, Suess, Steinberg, \&
  Sakurai}]{yamauchi2004differential}
Yamauchi, Y., Suess, S.~T., Steinberg, J.~T., \& Sakurai, T. 2004, Journal of
  Geophysical Research: Space Physics, 109

\bibitem[{Zank {et~al.}(2020)Zank, Nakanotani, Zhao, Adhikari, \&
  Kasper}]{zank2020origin}
Zank, G., Nakanotani, M., Zhao, L.-L., Adhikari, L., \& Kasper, J. 2020, The
  Astrophysical Journal, 903, 1

\end{thebibliography}

\end{document}